\documentclass[pdftex,12pt,a4paper]{article}
\usepackage[affil-it]{authblk}
\usepackage[margin=0.7in]{geometry}
\usepackage{multicol}
\usepackage{enumerate}
\usepackage[pdftex]{graphicx}
\usepackage[font={small,it}]{caption}
\usepackage[margin=8pt]{subfig}
\usepackage{multirow}
\usepackage{longtable}
\usepackage{float}
\usepackage{hyperref}	
\usepackage{lineno}

\begin{document}
\title{High purity 100 GeV electron identification with synchrotron radiation}
%

%
\author[12]{E.~Depero}
\author[12]{D.~Banerjee}
\author[10]{V.~Burtsev}
\author[10]{A.~Chumakov}
\author[12]{D.~Cooke}
\author[5]{A.~V.~Dermenev}
\author[9]{S.~V.~Donskov}
\author[6]{F.~Dubinin}
\author[10]{R.~R.~Dusaev}
\author[12]{S.~Emmenegger}
\author[4]{A.~Fabich}
\author[2]{V.~N.~Frolov}
\author[8]{A.~Gardikiotis}
\author[5]{S.~N.~Gninenko}
\author[1]{M.~H\"osgen}
\author[5]{A.~E.~Karneyeu}
\author[1]{B.~Ketzer}
\author[5]{M.~M.~Kirsanov}
\author[3]{I.~V.~Konorov}
\author[7]{V.~A.~Kramarenko}
\author[11]{S.~V.~Kuleshov}
\author[10]{V.~E.~Lyubovitskij}
\author[2]{V.~Lysan}
\author[2]{V.~A.~Matveev}
\author[9]{Yu.~V.~Mikhailov}
\author[2]{V.~V.~Myalkovskiy}
\author[2]{V.~D.~Peshekhonov\footnote{Deceased}}
\author[2]{D.~V.~Peshekhonov}
\author[9]{V.~A.~Polyakov}
\author[12]{B.~Radics}
\author[12]{A.~Rubbia}
\author[9]{V.~D.~Samoylenko}
\author[6]{V.~O.~Tikhomirov}
\author[5]{D.~A.~Tlisov}
\author[5]{A.~N.~Toropin}
\author[10]{B.~Vasilishin}
\author[11]{G.~Vasquez Arenas}
\author[11]{P.~Ulloa}
\author[12]{P.~Crivelli\footnote{Corresponding author, crivelli@phys.ethz.ch}}


\affil[1]{\it Universit\"at Bonn, Helmholtz-Institut f\"ur Strahlen-und Kernphysik, 53115 Bonn, Germany} 
\affil[2]{\it  Joint Institute for Nuclear Research, 141980 Dubna, Russia}
\affil[3]{\it   Technische Universit\"at M\"unchen, Physik Department, 85748 Garching, Germany}
\affil[4]{\it CERN, European Organization for Nuclear Research, CH-1211 Geneva, Switzerland}
\affil[5]{\it Institute for Nuclear Research, 117312 Moscow, Russia}
\affil[6]{\it P.N. Lebedev Physics Institute, Moscow, Russia, 119 991 Moscow, Russia}
\affil[7]{\it Skobeltsyn Institute of Nuclear Physics, Lomonosov Moscow State University, Moscow, Russia}
\affil[8]{\it Physics Department, University of Patras, Patras, Greece} 
\affil[9]{\it State Scientific Center of the Russian Federation Institute for High Energy Physics of National Research Center 'Kurchatov Institute' (IHEP), 142281 Protvino, Russia}
\affil[10]{\it Tomsk Polytechnic University, 634050 Tomsk, Russia}
\affil[11]{\it Universidad T\'{e}cnica Federico Santa Mar\'{i}a, 2390123 Valpara\'{i}so, Chile}
\affil[12]{\it ETH Zurich, Institute for Particle Physics, CH-8093 Zurich, Switzerland}

%
%
%
\vskip 0.25cm
\maketitle 
\bibliographystyle{utphys}
\clearpage
\begin{abstract}
In high energy experiments such as active beam dump searches for rare decays and missing energy events, the beam purity is a crucial parameter. In this paper we present a technique   to reject heavy charged particle contamination in the 100 GeV electron beam of the H4 beam line at CERN SPS. The method is based on the detection with BGO scintillators of the synchrotron radiation emitted by the electrons passing through a bending dipole magnet. A 100 GeV $\pi^-$ beam is used to test the method in the NA64 experiment resulting in a suppression factor of $10^{-5}$ while the efficiency for electron detection is $\sim$95\%. The spectra and the rejection factors are in very good agreement with the Monte Carlo simulation. The reported suppression factors are significantly better than previously achieved.
\end{abstract}

\section{Introduction}

Many high energy experiments require pure  electron beams. Despite the steady improvement of the beam lines, contamination below a level of few \% is very difficult to achieve.  An example is the NA64 experiment at CERN in which it is mandatory to suppress hadron and muon contamination in the electron beam since such particles  can generate irreducible background processes mimicking the experimental signature of a dark photon \cite{P348-proposal, Banerjee:2016tad}.
NA64 uses 100 GeV electrons from the H4 SPS beam line at CERN which is one of the best existing beam lines at this energy in terms of beam purity \cite{H4beamline}. 

The standard technique of detecting Cherenkov radiation to distinguish between electrons and heavy charged particles is very inefficient at energies of 100 GeV \cite{synchro-book}. Instead in NA64 we detect the synchrotron radiation, produced by the electrons passing through a dipole bending magnet, using BGO crystals located  downstream the magnet. NA64 is a fixed-target experiment at the CERN SPS \cite{Banerjee:2016tad} combining the active beam dump and missing energy techniques to search for rare events predicted in Dark Sector models \cite{Alexander:2016aln}. NA64 uses 100 GeV electrons from the H4 SPS beam line at CERN. The electron beam is dumped in an active target, an electromagnetic calorimeter (ECAL) made of lead and scintillators in a sandwich geometry of shashlik type (corresponding to about 40 radiation lengths) \cite{Banerjee:2016tad}. If Dark Sectors exist, Bremsstrahlung photons generated in the target could produce dark photons ($A'$) via kinetic mixing \cite{okun,galison,holdom}. Those could then decay in a pair of particles in the Dark Sector ($\chi$) which will escape the setup undetected. 
 
 The basic concept of the synchrotron radiation emitted from high energy electrons to reject heavy charged particles is to exploit the high suppression of the radiated power emitted by particles heavier than electrons passing through a magnetic field in order to discriminate them.  The use of this technique is not new and detection of electrons or positrons in electrons beams with momenta ranging from 30 to 50 GeV was reported earlier by \cite{synchro-exp,synchro-exp2,CRITTENDEN1989643} and the achieved suppression factor was around 1\%. 
However, here we present the results obtained with this method using the BGO detector employed in NA64. As will be shown, our results have a very good overall efficiency 95\% and 3 orders of magnitude better rejection factor than previously reported using a similar scheme \cite{synchro-exp2}.

A charged particle in a magnetic field moves in a circular motion
emitting photons along its trajectory due to the basic principles of
electrodynamics. Both quantum and classical theory of synchrotron
radiation (SR) are well understood \cite{synchro-book}. In the range of
interest for our experiment both treatments are equivalent and we can
therefore use the classical approximation for our calculations. The
total power $S$ emitted per unit length by a relativistic charged particle of
energy $E$ with mass $M$ and with bending radius $R$ in a magnetic
field $B$ perpendicular to its velocity is given by:
\begin{equation}
S = \frac{q^2c}{6\pi}\frac{1}{(Mc^2)^4 }\frac{E^4}{R^2}
\label{eqn:power-emitted}
\end{equation}
where $q$ is the charge of the particle and $c$ the speed of light,
Since the emission angle of the synchrotron photons is proportional to
the inverse of the Lorentz factor $\gamma$, the photons are emitted
tangentially to the particle trajectory.\par

The total emitted power has an inverse scaling to the fourth power of
the charged particle mass. Therefore, heavy charged particles emit
orders of magnitude less synchrotron radiation than light ones.  For
$\mu^{+/-}$ and $\pi^{+/-}$ with about 200 $e^-$ mass, one can estimate
that they radiate $\sim  10^{-9}$ times less than an electron. This
would be the case if the particles propagate in an ideal vacuum.
However, in a real experimental setup, vacuum windows, residual gas,
beam counters such as scintillators and trackers result in interactions
of the incoming particles with material. Therefore, the suppression factor when crossing materials is limited by the emission of secondary electrons with enough kinetic energy (several MeV) to leave a synchrotron-like signal in the detector. Although most of the energy transfer due to ionization for heavy
charged particles is only a few keV, rare high energy transfer is
possible. The distribution of such secondary electrons with
kinetic energy $T\gg I$, where $I$ is the mean excitation energy of the
atom/molecule, for a particle with velocity $\beta=v/c$ and charge $z$
passing through a material with atomic number $Z$, mass number $A$ and
thickness $dx$ is described by \cite{PDG}:
\begin{equation}
\frac{d^2N}{dTdx} = \frac{1}{2}K z^2 \frac{Z}{A}
\frac{1}{\beta^2}\frac{F(T)}{T^2}
\label{eqn:knock-on}
\end{equation}
The constant $K$ is defined as $K=4\pi N_A r_e^2 m_e c^2$ where N$_A$ is
the Avogadro's number, $r_e$ is the classical electron radius and $m_e$
the electron mass.
$F(T)$ is a spin-dependent factor, which in our case for $T \ll W_{max}$
is very close to unity. W$_{max}$ is the maximal energy transfer in a single collision to the
electron:
\begin{equation}
W_{max} = \frac{2m_e c^2 \beta^2 \gamma^2}{1+ 2\gamma m_e/M+(m_e/M)^2}
\end{equation}
For a $\pi^-$ at 100 GeV, $W_{max}$ is roughly 1 GeV which
covers completely the energy range where synchrotron radiation is
emitted. Eq.\ref{eqn:knock-on} is valid in the range $I\ll T \leq W_{max}$.

The total emitted power scales with the energy of the charged particle to the fourth power. In principle, this scaling would make it possible to correlate the total energy emitted via synchrotron radiation with the momentum of the particle. However, the broad total energy distribution of the synchrotron radiation would result in a very low
efficiency for rejecting 50 GeV electrons in a 100 GeV beam. Furthermore, experimental issues limit this approach. In fact, electrons interacting with the material or the residual gas can generate Bremsstrahlung photons that overlap with the synchrotron radiation spectrum and spoil the energy reconstruction capability. For this reason, in NA64 such an approach alone is not enough to achieve the sensitivity required by the experiment and therefore a tracking system is mandatory to reject low energy electrons \cite{MM}.\par

\section{Experimental setup}
\begin{figure}[h!]
  \centering
  \includegraphics[width=1.\textwidth]{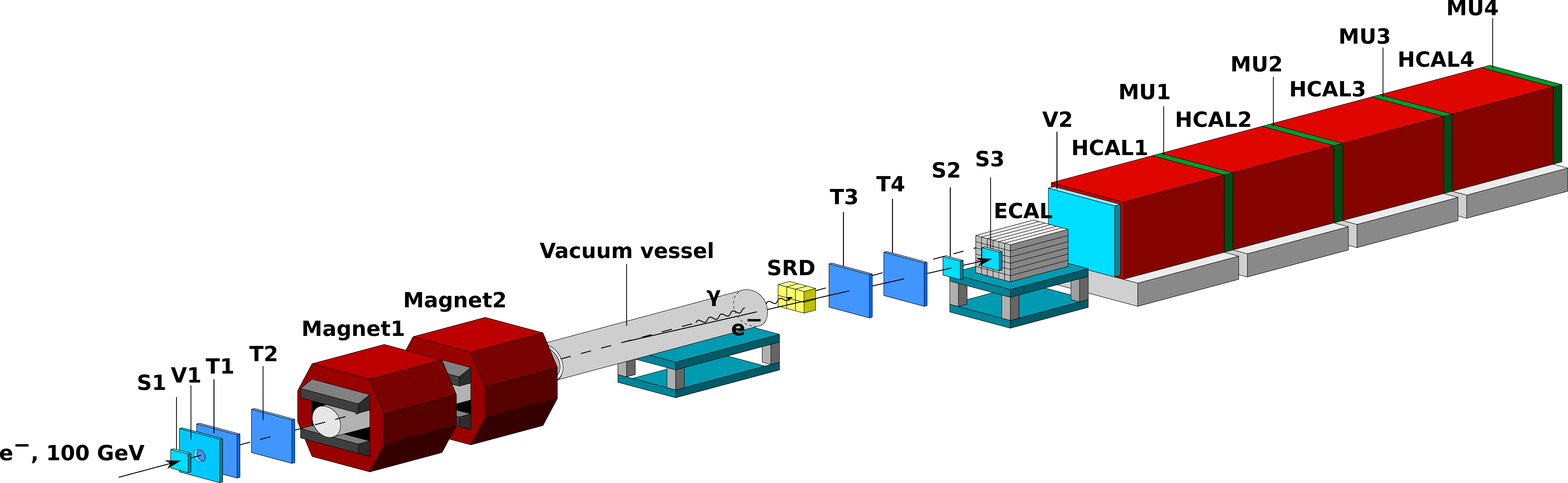}
  \caption{Drawing of the NA64 experimental setup used for these measurements (from \cite{Banerjee:2016tad}).}
  \label{fig:NA64Setup}
\end{figure}

\begin{figure}[h!]
  \centering
   \includegraphics[width=1.\textwidth]{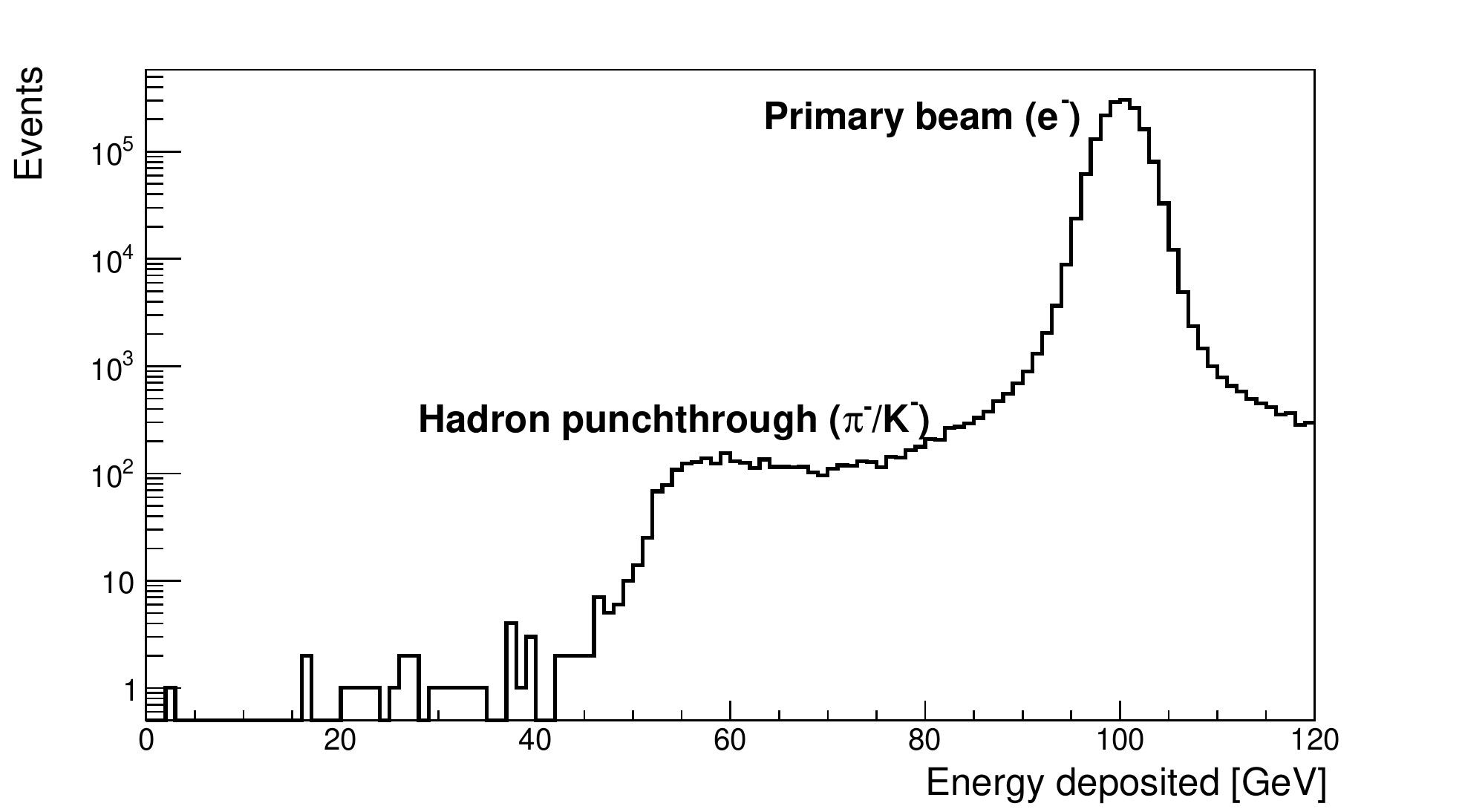}
  \caption{Measured spectra of the energy deposited in the ECAL for the 100 GeV electrons delivered by the H4 beam line.}
  \label{fig:ECAL_energy}
\end{figure}

The NA64 experimental setup \cite{Banerjee:2016tad} used for this measurement is shown in Fig.\ref{fig:NA64Setup}. As shown in Fig. \ref{fig:NA64Setup}, a set of veto counters (denoted as V2 in the drawing) and 4 hadronic calorimeters (HCAL) modules placed directly after the ECAL ensures the hermeticity required to search for the missing momentum with the aimed sensitivity. The experimental signature is thus an energy deposition below a given threshold (around 50 GeV in NA64) in the ECAL, no activity in V2 and no energy deposition in the HCAL. In Fig. \ref{fig:ECAL_energy} the energy deposition in the ECAL for a 100 GeV electron beam is shown. In addition to the expected peak at 100 GeV from electrons, one can see the presence of hadrons punching through the ECAL and the peak close to zero from minimum ionising particles (MIP).
Since hadrons and muons are a source of background for the experiment \cite{P348-proposal}, their contamination in the beam has to be suppressed at a level of $10^{-5}$. The synchrotron radiation detector (SRD) described in this paper was designed to serve this purpose. 

To reject low energy electrons and heavier charged particles, a synchrotron radiation detector (SRD) requires good energy resolution (see Eq. \ref{eqn:power-emitted}). Moreover, in order to work efficiently at high rates of $1\times 10^5$ particles/s good timing properties are necessary.
 
We chose Bi$_4$Ge$_3$O$_{12}$ (BGO) crystals the properties of which are a good compromise to fulfil these requirements. 
The detector consists of 8 hexagonal crystals with an external diameter of 55 mm and a length of 200 mm. The crystals are grouped into two modules. Each crystal is wrapped in Teflon tape for efficient light collection and it is glued to an ETL 9954 photomultipliers (PMT). The gain of the PMTs was set using a 100 GeV $\pi^-$ beam impinging directly on the SRD. The pions crossing the BGO have a mean energy deposition of about 60 MeV as expected for a MIP particle which is in the range of the SR radiation.

The BGO has a density of 7.13 g/cm$^3$ and because of the high atomic number of the bismuth component (Z=83) it has one of the largest probability per unit volume for photoelectric absorption of gamma rays \cite{Knoll}. The light yield of about 8500 photons/MeV coupled to the transportation losses and quantum efficiency of the PMT gives an energy resolution of about 17 \% (FWHM) at 1.27 MeV (measured with a $^{22}$Na radioactive source). All the crystals were tested using cosmic radiation and their decay time was measured to be about 300 ns in good agreement with the value quoted in the literature \cite{Knoll}. The BGO's signals are digitized by the MSADC system \cite{MSADC} which is connected to the DAQ of the
experiment. The DAQ  is downscaled version of COMPASS iFDAQ system \cite{daq}.

\section{Geant 4 simulation of SR spectra} 
Simulation of the expected SR signal was performed with the Geant 4 package\cite{geant4}.
The geometry of the NA64 experiment was coded in Geant 4, including the 200 $\mu$m mylar vacuum windows, the detailed composition of the trackers, scintillators and the residual gas was set at a level of $10^{-3}$ mBar as in the measurements. The BGO saturation was taken into account using Birks' law with the constants taken from \cite{birk}.

\begin{figure}[h!]
\centering
\includegraphics[width=1.\textwidth]{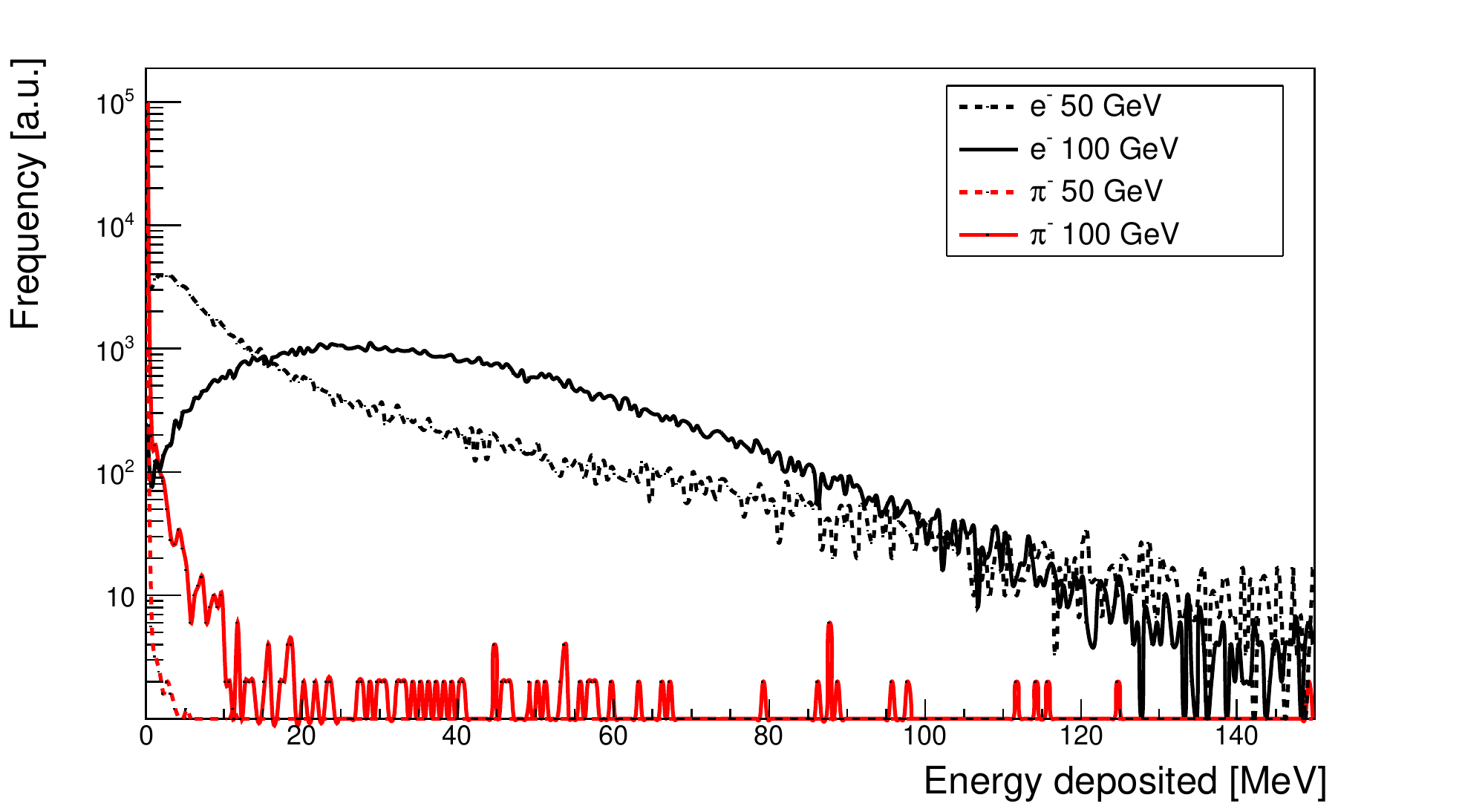}
\caption{Result of the Geant 4 simulation for the energy detected by the SR detector for 50/100 GeV e$^-$(black dashed/solid line) and 50/100 GeV $\pi^-$ (red dashed/solid line).}
\label{SRspectrum}
\end{figure}

The expected SR spectra for pions and electrons with energies of 50 GeV and 100 GeV are shown in Fig. \ref{SRspectrum}. The plot shows the expected dependence on the incoming electron energy in the emission spectra for the realistic experimental conditions.
Moreover, the comparison between the SR spectra of pions and electrons illustrates clearly the principle of this technique that allows to discriminate between them by requiring an energy threshold in the synchrotron detector. 
For pions, one can see that the probability of detecting an event with energy above 1 MeV (the threshold in our detector) is about $\sim 10^{-3}-10^{-4}$.
These SR-like signals originate from the interactions of the incoming pions with material which they ionise as predicted by Eq. \ref{eqn:knock-on}.
 \par 
Furthermore, Geant 4 reproduces the critical energy $E_c$ which divides the spectrum into two parts of equal power is:
\begin{equation}
E_c = \frac{3 \hbar c \gamma^3}{2R}
\end{equation}
with the reduced Plank constant $\hbar$ and the bending radius $R$. 
 For 100 GeV electrons in the  $B=1.7$ T bending field this corresponds to $E_c\sim$11.35 MeV. The expected mean energy of a synchrotron photon $E_m=E_c/\pi\simeq 3.6$ MeV is in very good agreement with simulation. The number of photons emitted per revolution in this energy range in the field of 7 T$\cdot$m is defined as:
\begin{equation}
N_\gamma = \frac{5 \pi \alpha}{\sqrt{3}}\gamma
\end{equation}
where $\alpha$ is the fine structure constant. 
By scaling this equation for the fraction of the circle where the particles are inside the magnetic field, one obtains a mean number of emitted photon of about 24.
The SRD geometrical acceptance is about one third,  thus one can estimate that the sum of deposited energy is approximately 29.35 MeV in good agreement with the results of the simulation as shown in Fig. \ref{SRspectrum}. 
 
\section{Experimental results}

\begin{figure}[h!]
  \centering
  \includegraphics[width=.9\textwidth]{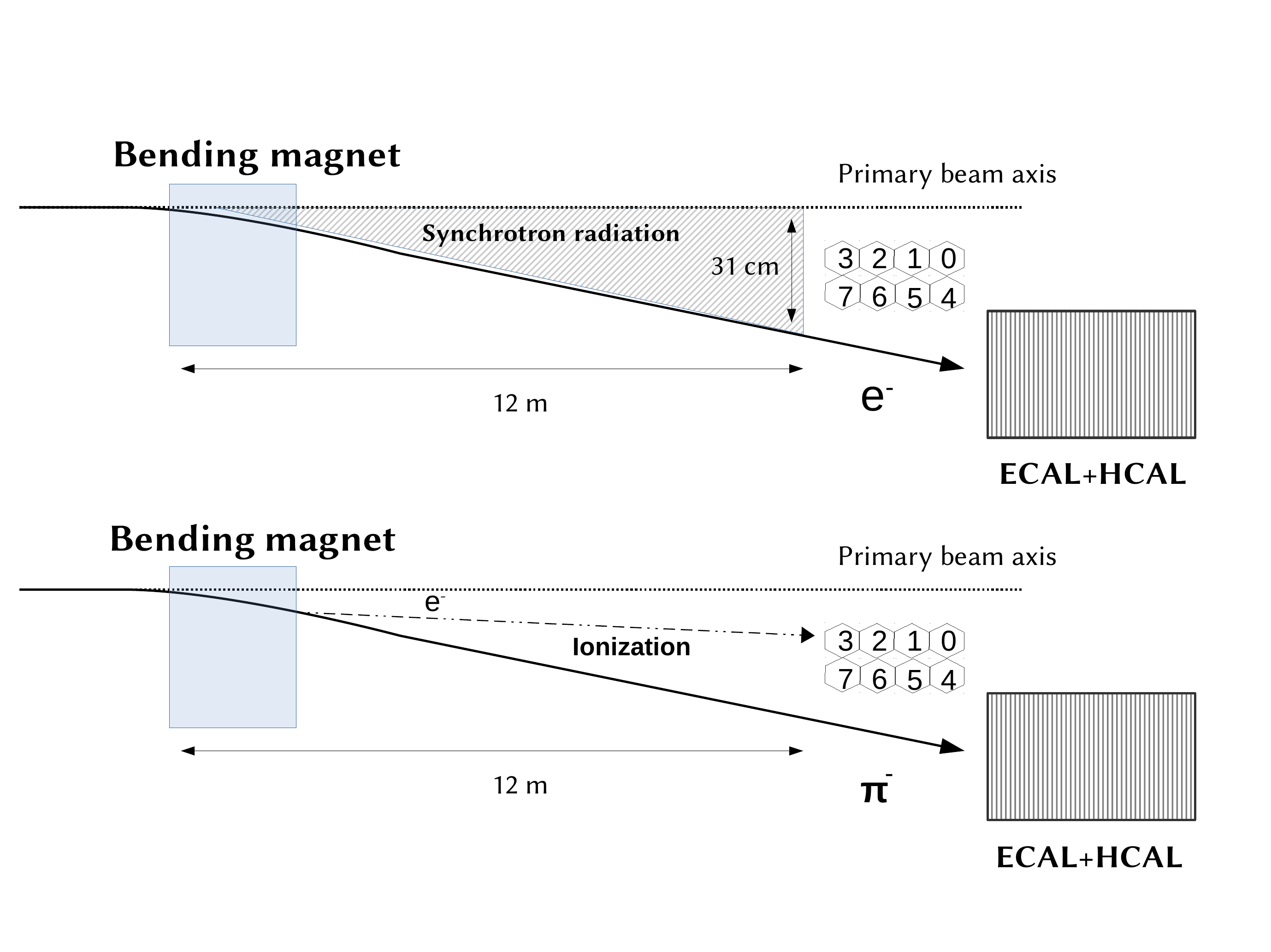}
  \caption{Geometry of the BGO crystals. Crystals 7,3 (SRD BGO) collect most of the synchrotron radiation spectrum. Crystals 4,0 (VETO BGO) on the other hand are effected only in case of a high energy event and are thus used as a veto. The remaining crystals serve as a shield for the SRD from backscattering particles coming from the ECAL. Top: illustration of event leaving a SR signal in the SRD. Bottom: illustration of a SR- like signal in the SRD for a knock-on electron produced by pions.}
\label{fig:newgeo}
\end{figure}

The SRD detector was tested during the NA64 test beam run in July 2016. The two BGO rows are parallel to the primary beam direction as shown in Fig.\ref{fig:newgeo}. The dipole magnets installed in series (see Fig. \ref{fig:NA64Setup}) produce a total integrated magnetic field of 7 T$\cdot$m \cite{Banerjee:2016tad} resulting in a nominal deviation for the incoming electrons at the SRD/ECAL positions of 31/34 cm from the undeflected beam axis. The two rows of BGO have about 9 cm distance from both the undeflected and the deflected beam axis (Fig.\ref{fig:newgeo}). This separation minimises the possibility for Bremsstrahlung photons and neutrals produced in upstream beam particle's interactions with material (vacuum windows, collimators, trackers...) and for particles in the beam halo (the typical beam diameter used in the experiment was 2 cm) to hit the SRD. In fact, such interactions result in the saturation of the SRD with a significant loss of efficiency due to the long decay time of the BGOs.


The two crystals facing the beam (labeled 3 and 7 in Fig. \ref{fig:newgeo}) detect most of the energy emitted by the synchrotron radiation. We will refer to those as SRD BGOs from now on. The remaining six crystals are used to detect events with high energy deposition in the SRD. In particular the last two crystals of each row (labeled 0 and 4 in Fig. \ref{fig:newgeo}) detect some energy only in the case of very energetic Bremsstrahlung events and thus can be used as a veto (see Fig.\ref{fig:newgeo}). The six crystals after the SRD BGOs act also as a shield from backscattering particles coming from the ECAL suppressing pions by an additional order of magnitude. Finally in this new geometry it is possible to use the coincidence of the two SRD BGO crystals to improve the tagging of synchrotron photons by rejecting knock-on electrons produced by incoming pions. In fact synchrotron radiation has a homogenous spectrum in the whole arc described by the primary and deflected beam and thus a signal is detected in both SRD BGOs. On the contrary, electrons generated by a $\pi^-$ undergoing hadron ionisation will mostly leave energy only in a single crystal as illustrated in Fig. \ref{fig:newgeo}. 
With the requirement of detecting in both SRD BGOs an energy deposition above a 1 MeV the suppression factor is improved up to a level of $10^{-5}$.

Data with a 100 GeV $\pi^-$ beam were taken to have a direct measurement of the suppression factor achievable for hadrons through synchrotron radiation measurements. The beam intensity was 5.3$\times 10^4$ particles per spill. The trigger was given by the coincidence of the three plastic scintillator counters (S1, S2 and S3 shown in Fig.\ref{fig:NA64Setup}). The additional requirement of an energy deposition below 60 GeV in the ECAL was applied in order to select an almost pure $\pi^-$ sample of $\sim 10^5$ collected events. The probability for electrons to punch through the 40 radiation length of the ECAL was estimated to be at a level of $10^{-12}$ \cite{Banerjee:2016tad}. 

For the 100 GeV electron beam run, a total of 220 spills were recorded with an intensity of 3.4$\times 10^5$/spill. 
The same trigger used in the pion run was used for the electron data.
In this case though, in order to reduce the pion contamination which is at a level of few \% and obtain a pure sample of electrons only events with a total energy deposition in ECAL + HCAL above 90 GeV but with less than 20 GeV energy in the HCAL were used.

\begin{figure}[h!]
	\centering
		\includegraphics[width=1.\textwidth]{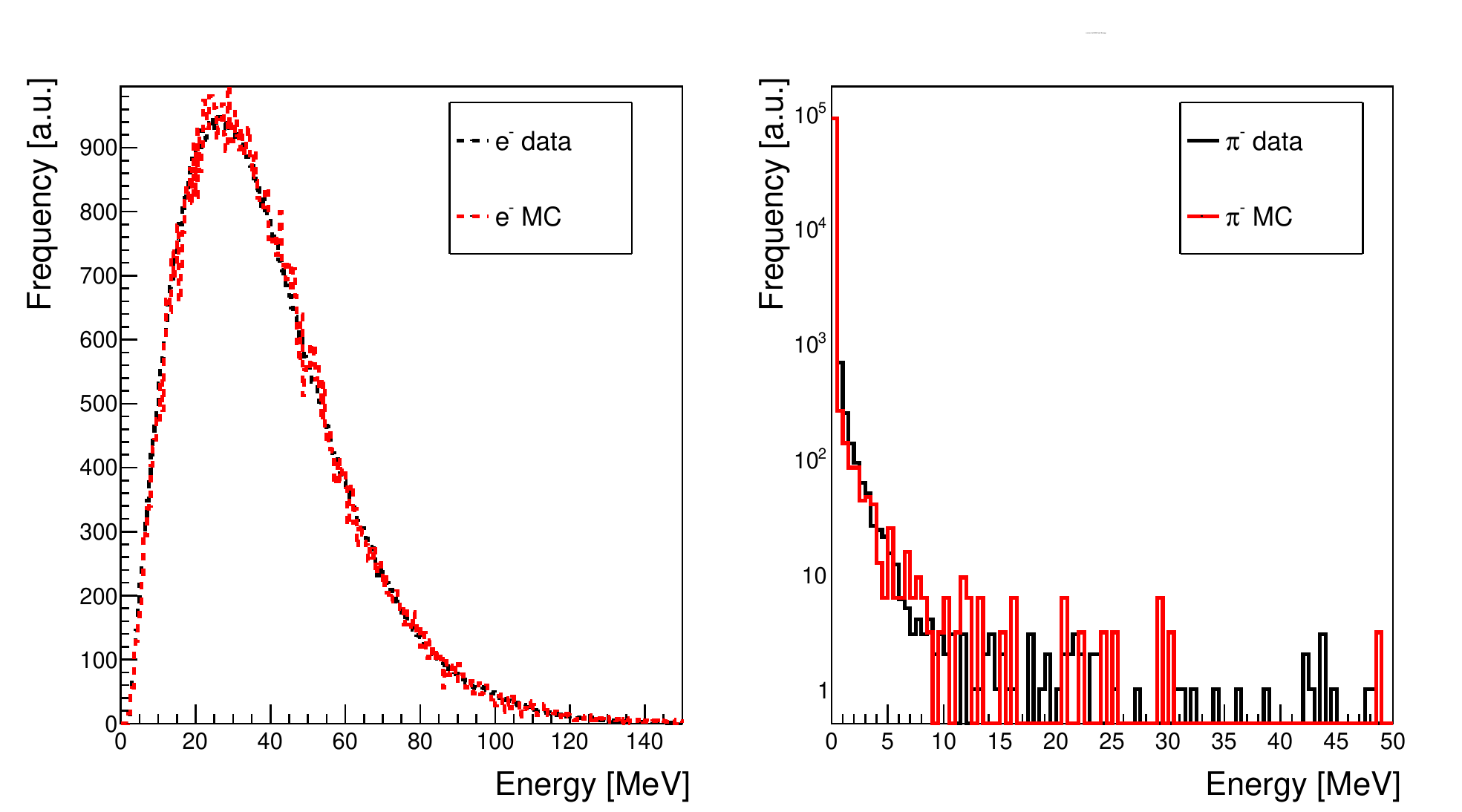}
	\caption{Comparison between data and simulation (MC) of the synchrotron radiation spectrum detected for 100 GeV electrons (left) and pions (right). }
	\label{fig:comp_spectra}
\end{figure}

The energy spectra recorded by the SRD BGO with electrons and pions are shown in Fig.\ref{fig:comp_spectra}. The SR spectra obtained with the electron beam are used to perform the BGO calibration by comparison with the simulation. With this method a very good agreement of data and MC is achieved (see Fig.\ref{fig:comp_spectra} plot on the left). As a cross check using the obtained calibration constants, the data from the pion beam impinging directly on the SRD are fit with a Landau distribution. The obtained peak position of 60 MeV is in good agreement with the prediction of the MC. 

Time coincidence of signals above the energy threshold of 1 MeV from both SRD counters is required and high energy Bremsstrahlung events are removed using the veto BGO.
The suppression of synchrotron radiation emission detected for pions compared to electrons is clearly visible by comparing the two plots. For the electron spectrum, a 1\% pileup beam events have been added to the simulation as predicted for the given spill intensity and with the known decay time of BGOs.  Both spectra are in very good agreement with the simulation.

\begin{figure}[h!]
  \centering
  \includegraphics[width=1.\textwidth]{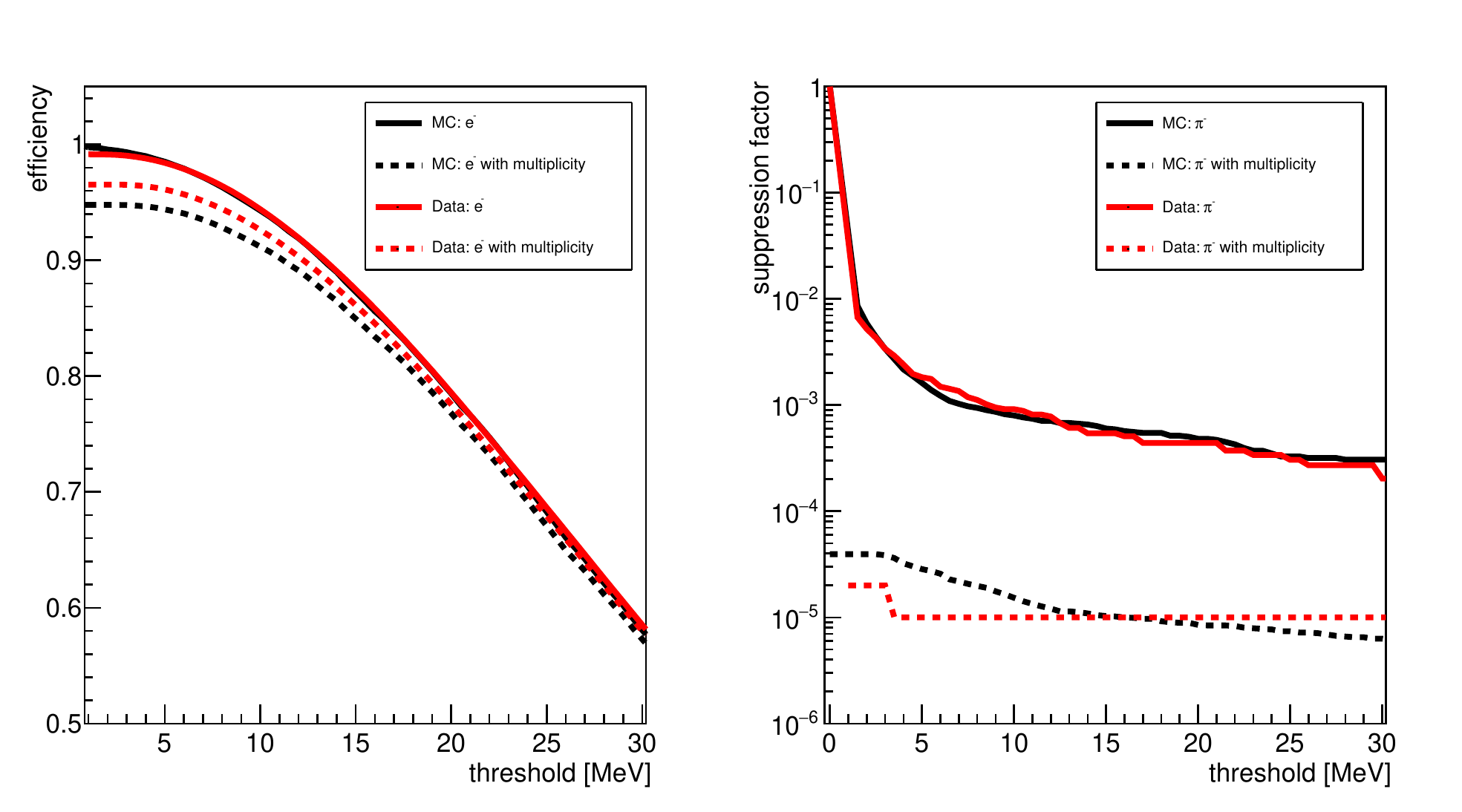}
  \caption{Left: Comparison between data and simulation (MC) for electrons of the efficiency as a function of threshold set on the total energy deposited in the SRD BGO and for the requirement that this is deposited in each single crystal (multiplicity). Right: Comparison between data and simulation for pions and electrons of the suppression factor as a function of the threshold set on the total energy deposited in the SRD BGO and for the multiplicity requirement.}
  \label{fig:sup_mult}
\end{figure}

The efficiency for the electrons and the suppression factor for the pions are plotted in Fig. \ref{fig:sup_mult} as a function of the threshold on the energy deposited in the SRD. We distinguish between two cases:
\begin{enumerate}
\item The threshold is set on the total energy deposited in the SRD.
\item Both SRD signals have to be in-time and above the threshold (multiplicity requirement).
\end{enumerate}
One can see that applying the criterion 2) the efficiency only decreases slightly compared to 1), while the suppression factor for pions is dramatically increased (by two orders of magnitude) with the requirement of having the two BGOs in coincidence.
This can be understood because the SR-like signal generated from secondary electron will leave a signal only in one of the two BGOs while the SR from electrons is spread out uniformly as explained above. 
This is also nicely evidenced by Tab. \ref{tab:hits} where the fraction of events with different hit multiplicity in the SRD BGO for both pion and electron runs are reported.

\begin{table}[h!]
\begin{center}
\begin{tabular}{cccc}
Events hit multiplicity  (\%) & 0 BGO  & 1 BGO & 2 BGOs\\
\hline
Pions & $98.77$ & $1.21$ & $1.4\times10^{-3}$  \\
Electrons & $2.4\times10^{-1}$  & $2.60$ & $97.37$ \\
\end{tabular}
\end{center}
\caption{Fraction of pion and electron events for different hit multiplicity in the SRD from the data.}
\label{tab:hits}

\end{table}

\section{Conclusions} 
In this paper we demonstrated that the detection of synchrotron radiation using BGO crystals is a very powerful method to tag 100 GeV electrons bending in a 7 T$\cdot$m magnetic field.  Discriminating on the total energy deposited in the SRD by setting a threshold of 10 MeV, the contamination of pions in the beam can be suppressed down to a level of $10^{-3}$ (see Fig. \ref{fig:sup_mult}).
Moreover, we have shown that exploiting the granularity of the detector one can suppress the signal from secondary electrons generated via hadron ionization resulting in a suppression of hadron contamination down to a level $10^{-5}$ maintaining an electron identification efficiency of $\sim$95\%. According to the Geant 4 simulation validated with our measurements a further improvement by an additional order of magnitude is expected by increasing the detector granularity. In order not to degrade the efficiency due to pileup effects at higher beam intensity of $1\times 10^6$ electrons/s as achievable in the H4 beam line at the CERN SPS, one should use faster detectors such as, e.g. a PbSc sampling calorimeter or LYSO crystals. These options will be tested by the NA64 collaboration in the near future. 

\section*{Ackowledgements} 
We gratefully acknowledge the support of the CERN management and staff  and the technical staffs of the participating institutions for their vital contributions.  This
work was supported by the HISKP, University of Bonn (Germany),  JINR  (Dubna),  MON  and  RAS  (Russia), the Russian Federation program ``Nauka'' (Contract No. 0.1764.GZB.2017), ETH Zurich and SNSF Grant No. 169133 (Switzerland),  and  grants  FONDECYT  1140471 and  1150792,  PIA/Ring  ACT1406  and  PIA/Basal  FB0821  CONICYT (Chile).  Part of the work on MC simulations was supported by the RSF grant 14-12-01430. 
We  thank the COMPASS  DAQ  group  and  the  Institute  for  Hadronic Structure  and  Fundamental  Symmetries  of  TU  Munich for the technical support.

\end{document}